\documentclass[%
aip,amsmath,amssymb, reprint,
]{revtex4-1}
\usepackage{enumerate}

\usepackage{graphicx}
\usepackage{dcolumn}
\usepackage{bm}
\usepackage[ruled,vlined]{algorithm2e}
\usepackage[utf8]{inputenc}
\usepackage[T1]{fontenc}
\usepackage{mathptmx}
\usepackage{listings}
\usepackage{tikz}
\usepackage{soul}
\usetikzlibrary{arrows}
\usetikzlibrary{trees}
\usepackage{midfloat}
\usepackage{cancel}
\usetikzlibrary{decorations.pathmorphing}
\usetikzlibrary{decorations.markings}
\usetikzlibrary{automata,positioning}
\usepackage{float}
\usepackage{caption}
\usepackage{simplewick}
\usepackage{csquotes}
\usepackage{graphicx}
\usepackage{subcaption}
\usepackage{multirow}
\usepackage{amsmath}
\usepackage{cleveref}
\usepackage{tabularx}
\usepackage{natbib}
\usepackage{braket}
\usepackage{xcolor}
\usepackage{xcolor}
\definecolor{codegreen}{rgb}{0,0.6,0}
\definecolor{codegray}{rgb}{0.5,0.5,0.5}
\definecolor{codepurple}{rgb}{0.58,0,0.82}
\definecolor{backcolour}{rgb}{0.95,0.95,0.92}
\lstdefinestyle{mystyle}{
    backgroundcolor=\color{backcolour},   
    commentstyle=\color{codegreen},
    keywordstyle=\color{magenta},
    numberstyle=\tiny\color{codegray},
    stringstyle=\color{codepurple},
    basicstyle=\ttfamily\footnotesize,
    breakatwhitespace=false,         
    breaklines=true,                 
    captionpos=b,                    
    keepspaces=true,                 
    numbers=left,                    
    numbersep=5pt,                  
    showspaces=false,                
    showstringspaces=false,
    showtabs=false,                  
    tabsize=2
}

\begin{document}

\title{Advancing Practical Quantum Embedding Simulations via Operator Commutativity Based State Preparation for Complex Chemical Systems}
\author{Dibyendu Mondal}\affiliation{Department of Chemistry, \\ Indian Institute of Technology Bombay, \\ Powai, Mumbai 400076, India}
\author{Ashish Kumar Patra}\affiliation{Qclairvoyance Quantum Labs, Secunderabad, TG 500094, India}
\author{Rahul Maitra}\email{rmaitra@chem.iitb.ac.in}\affiliation{Department of Chemistry, \\ Indian Institute of Technology Bombay, \\ Powai, Mumbai 400076, India} 
\affiliation{Centre for Quantum Information Computing Science \& Technology, \\ Indian Institute of Technology Bombay, \\ Powai, Mumbai 400076, India}

\begin{abstract}
Determining the exponentially scaled ground state wavefunction and the associated 
molecular properties remains one of the central challenges in quantum chemistry. 
Hybrid quantum-classical algorithms implemented on quantum computers offer a 
promising route toward addressing this problem. However, despite several successful 
demonstrations on small molecular systems, accurate simulations of large and 
chemically realistic molecules remain difficult due to the limited capability of 
noisy intermediate scale quantum (NISQ) hardware. To bypass the limitations of
NISQ devices, while simultaneously retaining the accuracy of the ground state 
energy estimations, we propose a dynamic ansatz construction strategy based on operator 
commutativity and energy driven screening within density matrix 
embedding theory (DMET) framework. The partitioning of the full system allows us to dynamically construct the 
ansatz over individual embedded subsystems, allowing each embedding problem be solved 
individually to a desired accuracy. 
The embedding Hamiltonian is updated in a self-consistent manner with
dynamically formulated wavefunction, and their coupled optimization
leads to accurate and efficient description of the overall system.
To assess the performance of this approach, we apply it to several molecular systems and chemical 
processes with up 
to 144 qubits. These simulations require at most 20 
qubits at a time and demonstrate improved accuracy and significantly reduced quantum 
gate requirements compared with conventional ansatze. We further investigate the impact of various 
fragmentation strategies and demonstrate the adaptability of our approach at
each step of the DMET self-consistency cycle that leads to significantly improved accuracy for  
strongly correlated system.

\end{abstract}
\maketitle
\section{Introduction}
Accurate treatment of electronic correlation remains one of the central challenges in 
quantum chemistry. In classical many-body electronic structure methods, an exact 
representation of the many electron wavefunction requires an exponentially growing number 
of determinants, which becomes computationally prohibitive for classical computers beyond 
a certain problem size\cite{expo,expo1}. Quantum computing, through the principles of quantum 
superposition and entanglement, provides a promising route to represent such 
exponentially large Hilbert space using only a polynomial number of qubits. Nevertheless, 
despite this theoretical advantage, the simulation of large and realistic chemical 
systems is still beyond the capabilities of present day quantum hardware because of the 
limited number of fault-tolerant qubits and restricted gate fidelities\cite{Preskill,circuit_depth,circuit2,circuit3}. In the current 
noisy intermediate scale quantum (NISQ) era, hybrid quantum-classical approaches such as 
the variational quantum eigensolver (VQE)\cite{Peruzzo_2014} have demonstrated encouraging results for small 
molecular systems. However, accurate simulation of realistic chemical systems would 
require hundreds to thousands of qubits along with significantly improved gate 
fidelities, which remain beyond the reach of existing quantum devices.

Within the VQE framework, strong correlation effects in small to moderately sized molecular 
systems can be captured by preparing a parameterized trial state or ansatz, \cite{ucc_review,Tilly_2022}
($\ket{\psi(\theta)}=U(\theta)\ket{\phi_{ref}}$) , starting from a suitably chosen and easy
to prepare reference state. The parameters of the ansatz are iteratively optimized using a classical routine until the expectation value of the molecular Hamiltonian 
($\hat{H}$) reaches convergence.

\begin{eqnarray}
    E(\theta)=\bra{\psi(\theta)}\hat{H}\ket{\psi(\theta)}
\end{eqnarray}
According to the Rayleigh-Ritz variational principle, the optimized energy obtained from 
this procedure provides an upper bound to the true ground state energy of the system. 
Consequently, both the accuracy of the computed ground state energy and the associated 
computational cost depend strongly on the choice of the trial ansatz. Considerable efforts 
have been devoted to design compact and chemically accurate ansatze, in the unitary 
coupled cluster (UCC) framework, which are capable of capturing strong correlation effects 
with manageable circuit depths
\cite{uccgsd,adapt,uiccsdn,COMPASS,compact,rbm,surge,energy_sort,compass_pro,rbm2}. However, most of these approaches have primarily been 
demonstrated for molecular systems of limited size. 

In realistic large chemical systems, electronic structure typically involves a mixture of 
strong and weak correlation effects. If the strongly correlated component can be 
efficiently treated within the quantum computing framework, it becomes possible to capture 
the essential chemistry of the system in a computationally efficient manner. In this 
context, several classical embedding and fragmentation techniques (density matrix 
embedding theory (DMET)\cite{dmet_initial,dmet_initial2,prac_dmet}, dynamical mean-field theory (DMFT)\cite{dmft1,dmft2}, localized active space self-consistent field method (LASSCF)\cite{lasscf,lasscf2}
fragment molecular orbital 
(FMO) theory\cite{fmo1,fmo2,fmo3}, density functional theory (DFT) embedding\cite{dft1,dft2}, bootstrap embedding\cite{boot,boot1,boot2}, tensor network approaches\cite{tensor1,tensor2}, and 
perturbative methods\cite{perturb}) have been developed, where a large system is partitioned into 
smaller subsystems to treat correlation more effectively. Despite their success, the 
accuracy of these techniques often depends critically on the fragment size. As the 
fragment becomes larger, highly accurate classical methods for treating strong correlation 
such as full configuration interaction (FCI), coupled cluster (CC) methods become computationally 
prohibitive. This limitation motivates the integration of efficient quantum computing algorithms 
to treat the strongly correlated fragments more effectively\cite{Li-dmet}.

To address these challenges, we adopt a hybrid strategy in which a large chemical system is first 
partitioned into smaller fragments at the mean-field level within the classical DMET framework 
\cite{dmet_initial,prac_dmet}. Subsequently, the localized electronic correlation 
within each fragment is captured  by the construction of a dynamic ansatz within the UCC framework 
\cite{ucc_review,COMPASS}. This ansatz is generated through a novel operator commutativity and energy 
screening procedure, where higher-rank terms are tensor decomposed into effective two-body 
cluster operators and `scatterers' to achieve a balanced description of many-body correlation effects, as 
introduced previously by some of the current authors\cite{iccsdn,COMPASS,uiccsdn,compass_pro}. The choice of these scatterers is guided by their 
non-commutativity with the cluster operators and their energetic relevance. This methodology, termed
COMmutativity Pre-screened Automated Selection of Scatterers (COMPASS)\cite{COMPASS}, has been shown to dynamically 
balance accuracy with quantum efficiency. However, in the DMET framework which provides a practical 
platform for studying large chemical systems by decomposing them 
into manageable fragments that are compatible with the limitations of current NISQ hardware, the 
adaptation of COMPASS is far form being trivial. This is primarily due to the non-trivial
coupling between the embedding Hamiltonians and the dynamically constructed ansatz for each subsystem. 
This complexity arises because within DMET, the consistency between the high-level correlated wavefunction 
and the low-level mean-field description is enforced through the introduction of a global chemical 
potential. This chemical potential is incorporated into each embedded Hamiltonian and iteratively 
optimized such that the total particle number obtained from all fragmented subsystem's high-level 
wavefunction matches 
that of the mean-field wavefunction of the full system.
Thus during this self-consistent optimization 
cycle, the dynamic high-level correlated wavefunction for each fragment must be reconstructed 
at every step with the updated chemical potential to ensure a consistent and balanced 
description across all subsystems. Thus in our proposed scheme the dynamic nature is not only at the 
level of ansatz
preparation for embedding subsystems, but also the embedding Hamiltonians get modified at each step 
requiring alternative optimization cycle for ansatze and the Hamiltonians. This formalism is 
abbreviated as DMET-COMPASS.

The rest of the manuscript is organized as follows: in Sec.\ref{dmet_intro} , we begin with an overview of the DMET framework, followed by the construction of high-level wavefunctions on a quantum computer in Sec.\ref{ansatz_intro}. The efficient dynamic ansatz construction strategy is discussed in Sec.\ref{dynamic_compass}, followed by a detailed description of the hybrid quantum-classical embedding framework in Sec.\ref{method}.
To validate the proposed approach, we present numerical results in Sec.\ref{results} for a range of molecular 
systems, including the circular $C_{10}$ system, multiple conformers of L-glucose, and 
the reaction energy profile of the (4+2) cycloaddition between cyclopentadiene and  methyl vinyl ketone. These systems require 100, 144, and 124 qubits, respectively, for full
system simulations on a quantum computer within the STO-3G basis. By employing an atom-wise partitioning strategy within the DMET framework, the effective problem size is 
reduced to at most 20 qubits. In Sec. \ref{frag_tech} , we systematically analyze the effects of various fragmentation sizes with strongly correlated linear $H_{12}$ system, and demonstrate 
how the size of the embedding problem critically impacts the performance of DMET-COMPASS. 
Finally in Sec. \ref{ansatz_adaptability}, we demonstrate the adaptability of the ansatz with the self-consistent iterative evolution of the embedding Hamiltonian, and show that the coupled optimization of the high-level wavefunction of the embedded 
subsystems along with the recursive update of the embedding Hamiltonian leads to 
improved description of the strongly correlated molecular systems. 

\section{Theory}
\subsection{Density Matrix Embedding Framework}\label{dmet_intro}
In the quantum computing paradigm, each spinorbital is usually mapped directly onto a 
qubit. Consequently, simulating a large chemical system demands an enormous 
number of fault-tolerant qubits, far exceeding the capabilities of present-day 
quantum hardware. To make realistic quantum chemical simulations feasible, it is 
therefore essential to minimize qubit requirements. 

One promising strategy is the use of quantum embedding methods, among which 
DMET\cite{dmet_initial,dmet_initial2}, is particularly powerful. DMET reformulates the 
original quantum problem into a reduced system consisting of a fragment, its 
entangled bath, and the surrounding pure environment. The central idea is to 
compress the Hilbert space by applying a Schmidt decomposition to the 
wavefunction. Concretely, the Hilbert space is partitioned into two orthogonal 
subspaces: the fragment
$A$ with dimension $L_A$ and environment $B$ with dimension $L_B$, ($L_B>L_A$). The full quantum state in the product basis 
{$\ket{A_i}\ket{B_j}$} which spans a dimension of $(L_A \times L_B)$, can then be expressed as:
\begin{eqnarray}
    \ket{\psi}=\sum_i^{L_A} \sum_j^{L_B} C_{ij}\ket{A_i}\ket{B_j}
\end{eqnarray}
where $C_{ij}$ is the coefficient tensor.
However, the effective size of the problem can be significantly reduced by explicitly accounting for the entanglement between the fragment and the environment. In particular, the quantum state $\ket{\psi}$ can be 
expressed in a rotated basis as 
\begin{equation}
    \ket{\psi}=\sum_{X}^{L_A}\omega_{X}\ket{\tilde{A}_{X}}\ket{\tilde{B}_{X}}
\end{equation}
which corresponds to the Schmidt decomposition of bipartite states. This decomposition 
naturally partitions the environment into two distinct parts: a set of at most $L_A$ 
bath states entangled with the fragment, and the remaining states that are completely 
disentangled. On this basis, the embedding Hamiltonian can be constructed by 
projecting the full Hamiltonian ($\hat{H}$) into the reduced subspace defined by the 
fragment and its associated bath. 
This is achieved through the relation $\hat{H}_{emb}=\hat{P}\hat{H}\hat{P}$ with the projector $\hat{P}$ defined as:
\begin{eqnarray}
    \hat{P}=\sum_{XY}\ket{\tilde{A}_X \tilde{B}_{Y}}\bra{\tilde{A}_X \tilde{B}_{Y}}
\end{eqnarray}
which is the projection onto at most $2L_A$ dimensional subspace, 
significantly smaller than the full
space. Although the constructed subspace is formally exact and demonstrates that the full 
system can, in principle, be represented by a fragment coupled to its entangled bath,
it remains impractical in practice. This is due to the fact that it relies on the prior 
knowledge of the exact ground state wavefunction and its corresponding Schmidt decomposition
which makes it somewhat impractical. In practice
DMET employs a mean-field approximation to the ground state wavefunction of the full system and 
generates the fragment and bath orbitals which together formed partitioned 
subsystem ($\xi$). 
Within each embedded fragment-bath space, the mean-field state ($\ket{\phi_{ref}^\xi}$) serves as a 
reference, and electron correlation may be incorporated through high-level correlation 
calculation using a suitable parametrized quantum circuit ($\ket{\psi^\xi(\theta_{opt})}$).

Since each subsystem is treated independently, the redistribution of electrons between fragment 
and bath orbitals during the DMET procedure can lead to an inconsistency in the electrons 
count. 
As a result, the total number of electrons obtained by summing over all fragments may deviate 
from that of the full system.
To enforce particle number consistency, a global chemical potential ($\mu_{gl}$) is 
introduced, which modifies the embedded Hamiltonian ($\hat{H}_{emb}^\xi$) as
\begin{eqnarray}\label{emb-hamiltonian}
    \hat{H}_{emb}^\xi(\mu_{gl})\leftarrow \hat{H}_{emb}^\xi-\mu_{gl}\sum_p^{L_A}a_p^{\dagger}a_p.
\end{eqnarray}

After obtaining the correlated high-level wavefunction for each embedded Hamiltonian ( $\hat{H}_{emb}^\xi(\mu_{gl})$), the one- (1-RDM: $D_{pq}^\xi=\bra{\psi^\xi(\theta_{opt})}a_p^{\dagger}a_q\ket{\psi^\xi(\theta_{opt})}$) and two-particle reduced density matrices (2-RDM: $P_{pqrs}^\xi=\bra{\psi^\xi(\theta_{opt})}a_p^{\dagger}a_q^{\dagger}a_ra_s\ket{\psi^\xi(\theta_{opt})}$) are calculated with the high level wavefunction.
These reduced density matrices are then used to evaluate the energy of each embedded subsystem:
\begin{eqnarray}
    E^\xi = \sum_p^{L_A} \Big(\sum_{q}^{L_A+L_B} \frac{t_{pq}+\tilde{h}_{pq}}{2} D_{pq}^\xi
    +\frac{1}{2}\sum_{qrs}^{L_A+L_B}h_{pqrs}P_{pqrs}^\xi \Big )
\end{eqnarray}
where $\tilde{h}_{pq}=(t_{pq}+\sum_{mn}^{L}(h_{pqmn}-h_{pnqm})D_{mn}^{env})$ 
represents the rotated one-electron integral which also considers the effect of the 
unentangled electrons. Finally the total energy of the entire system is calculated
by summing up all the fragment energy.
\begin{eqnarray}
    E_{total} = \sum_\xi E^\xi + E_{nuc}
\end{eqnarray}
Here $E_{nuc}$ is the nuclear repulsion energy of the entire system. 
In practice, this procedure is carried out self-consistently. Starting from an initial guess for the 
global chemical potential, the high-level wavefunction and corresponding electron number are computed for 
each subsystem. The chemical potential is then iteratively updated until the deviation between the total 
electron count (summed over all fragments) and that of the full system falls below a predefined 
threshold. Importantly, at each iteration, the embedded Hamiltonians are updated and the high-level 
wavefunctions are recomputed to ensure overall consistency. 

\subsection{Correlated High-level Wavefunction Generation on Quantum Computer}\label{ansatz_intro}
The correlated high-level wavefunctions for 
each embedded subsystem can, in principle, be obtained using accurate classical electronic structure 
methods. However, the overall accuracy of energy strongly depends on the size of the fragment: 
as the embedded subsystem becomes larger, the computational cost of reliable classical approaches grows 
rapidly, often rendering them impractical. In such cases, quantum computing algorithms provide a 
promising alternative, offering a more scalable route to accurately capture correlation effects in large 
embedded spaces. As discussed earlier, given the limitations of current quantum hardware, hybrid quantum-classical approaches such as VQE are more suitable for obtaining correlated wavefunctions. The accuracy and computational efficiency of VQE are largely determined by the choice of trial ansatz. Within the UCC framework, the trial ansatz for each embedded subsystem is constructed by applying anti-Hermitian excitation operators on top of the the mean-field reference state of the embedded subsystem.
\begin{eqnarray}
    \ket{\psi^\xi(\theta)}=e^{\tau^\xi(\theta)}\ket{\phi_{ref}^\xi}
\end{eqnarray}
Here $\tau^\xi(\theta)=T^\xi(\theta)-T^{\xi\dagger}(\theta)$ is the anti-hermitian cluster
operators and $\ket{\phi_{ref}^\xi}$ is the mean field state of embedded system $\xi$.
The general practice of truncating the cluster operators is 
up to the double excitation and thus UCC ansatz with single and double 
excitation operators provide a balance between accuracy and resource efficiency.
Although in the strong correlated regions higher order excitation effects which 
requires large number of quantum resources are
very important to capture the correlation effects effectively. 
Alternative strategy is required to capture such important phenomenon.

\subsection{Dynamic Ansatz Construction for the High-Level Wavefunction Generation}\label{dynamic_compass}
Direct inclusion of higher order excitation operators demands high quantum resources
although those effects are supremely important to generate accurate high-level 
wavefunction for each embedded system. To bypass this problem we have adopted a 
dynamic ansatz construction strategy where in addition to single and double excitation
operators, we have included a class of two-body generalized operators to capture
the higher order correlation effects implicitly. Therefore through our strategy one can 
generate important higher order effects (localized within each embedded subsystem) through lower body parametrization without incurring additional quantum resources. 

\subsubsection{A Class of Generalized Operators: Scattering Operators}
In the case of generalized operators excitation can occur from occupied to occupied or 
from unoccupied to unoccupied orbitals with respect to mean field reference.
The class of generalized operators that we have used has a specific structure.
It is two-body operators with effective excitation rank one. In details there is an 
excitation from occupied to 
unoccupied orbitals of mean field reference and a occupied to occupied or 
unoccupied to unoccupied scattering. Essentially there is a hole/particle 
destruction operators. Based on the type of destruction operator they are
classified as $S_h^\xi$ or $S_p^\xi$.
\begin{eqnarray}
    S_h^\xi = \theta_{ij}^{am}\{a_a^{\dagger}a_m^{\dagger}a_ja_i\} ;
    S_p^\xi = \theta_{ie}^{ab}\{a_a^{\dagger}a_b^{\dagger}a_ea_i\}
\end{eqnarray}
Here $i,j,..m$ and $a,b,...e$ denote the occupied and unoccupied orbital
indices with respect to the mean field reference ($\ket{\phi_{ref}^\xi}$) of embedded subsystem $\xi$. The introduction of scatterers essentially enables us to write every term of
three or higher rank excitation operators through tensor-decomposition of
two-body terms:
($(S^\xi)_{ij}^{am}(T^\xi)_{mk}^{bc}\longrightarrow (T^\xi)_{ijk}^{abc} \longleftarrow (S^\xi)_{ie}^{ab}(T^\xi)_{jk}^{ec}$) which are localized in $\xi$. 
The orbitals ($m \in \xi$ in $S_h^\xi$ and $e \in \xi$ in $S_p^\xi$) through which they 
contract with excitation operators are known as contractible
set of orbitals (CSOs). It is very important to note that the direct action 
of scatterers on the mean field reference is nilpotent $S^\xi\ket{\phi_{ref}^\xi}=0$. 
Thus the excitation operators and scatterers are chosen to be non-commutative partners 
to each other such that their combined effective action over the reference is nontrivial.
The anti-hermitian counterpart of higher order effects are generated through this 
non-commutativity:
\begin{eqnarray}
    [\sigma^\xi,\tau_2^\xi]\rightarrow \tau_3^\xi ; [\sigma^\xi,[\sigma^\xi,\tau_2^\xi]]\rightarrow\tau_4^\xi ...
\end{eqnarray}
Here $\sigma^\xi=S^\xi-(S^\xi)^\dagger$ and $\tau^\xi$ represent the
anti-hermitian scatterers and 
excitation operators of embedded subsystem $\xi$ respectively. 
The disentangled product of $exp(\sigma^\xi)$ and $exp(\tau^\xi)$ as they appear in an 
interwoven 
manner in our ansatz (see next section) introduces the exponential of effective higher rank 
excitation operator through the commutator. 
As the rank of commutators grows, the corresponding excitation ranks increase accordingly. 
While, in principle, higher-order effects can be constructed solely from rank-two operators, 
incorporating the full set of one and two-body excitation operators along with two-body 
scattering terms demands substantial quantum resources, exceeding the limits of current quantum 
hardware. Moreover, the realization of higher-order contributions is highly sensitive to the 
ordering of the chosen excitation and scattering operators. Therefore, an efficient and 
systematic strategy is essential to selectively screen the relevant cluster and scattering 
operators, as well as to determine their optimal ordering. We would present such a
strategy in the next subsection.

\subsubsection{COMPASS - a Dynamic Ansatz Tailoring Strategy for Embedded Subsystem}
The selection of dominant excitation operators and scatterers as well as their 
ordering in the trial ansatz play a very crucial role to determine accurate 
high-level wavefunction for each embedded subsystem. 
More importantly, the ordering of the operators and/or their structural orientation
may vary for each embedded subsystem, requiring for local optimization and construction
of the ansatz for each.
To address this, in the first step of our 
ansatz construction scheme, we systematically prune the relevant two-body excitation operators 
through a parallel single-parameter optimization procedure. Starting from the mean-field 
reference state ($\ket{\phi_{ref}^\xi}$) of the embedded system $\xi$, we construct a set of 
trial circuits by applying each two-body excitation operator individually. Each of these single 
parameter circuits is then independently optimized to obtain the corresponding minimum energy, 
enabling the identification of the most significant operators. 
\begin{eqnarray}\label{one_par}
    E_I^\xi=\bra{\phi_{ref}^\xi}e^{-\tau_I^\xi(\theta_I)}\hat{H}_{emb}^\xi(\mu_{gl})e^{\tau_I^\xi(\theta_I)}\ket{\phi_{ref}^\xi}
\end{eqnarray}
Here $I$ represents combined hole-particle indices of two-body excitation operators of embedded subsystem $\xi$.
Only those two-body excitation operators are included for which the energy 
difference $\Delta E_I^\xi=(E_{ref}^\xi-E_I^\xi)$ with the mean field reference 
state is greater than a predefined threshold $\Delta E_I^\xi >\epsilon_1$.
The selected two-body excitation operators are placed into several operator
blocks ($\alpha,\beta,\gamma...\in \xi$) in a descending order of their energy contribution 
on the reference state. This implies that energetically most dominating two-body operator
is allowed to act on the reference state directly. This means that among the 
selected two-body excitation operators
if $\Delta E_I^\xi...>\Delta E_J^\xi...>\Delta E_K^\xi...$, the corresponding operators are 
placed in operator block
$\alpha$, $\beta$ and $\gamma$ respectively where $...<\gamma...<\beta...<\alpha...
<1$. After this step the ansatz structure is following 
\begin{eqnarray}
    \ket{\psi^\xi(\theta)}=...U_{\gamma}^\xi...U_{\beta}^\xi...U_{\alpha}^\xi...\ket{\phi_{ref}^\xi}
\end{eqnarray}
where $U_{\alpha}^\xi=[e^{\tau_I^\xi(\theta_I)}]_{\alpha}$.
In the subsequent step, each operator block is further enriched by incorporating suitable 
scattering operators. For every selected two-body excitation operator within a given block, we 
sequentially examine the non-commutativity of potential scatterers based on the presence of 
common CSOs with that excitation operator. A scatterer sharing common CSOs effectively 
generates higher-order excitation effects. We then assess the significance of these generated 
triple-excitation contributions by evaluating their impact on the energy. This is achieved 
through the optimization of a corresponding two-parameter quantum circuit.
\begin{eqnarray}\label{two_par}
    E_{I\nu}^\xi=\bra{\phi_{ref}^\xi}e^{-\tau_I^\xi(\theta_I)}e^{-\sigma_{\nu}^\xi(\theta_\nu)}
    \hat{H}_{emb}^\xi(\mu_{gl})\\ \nonumber
    e^{\sigma_{\nu}^\xi(\theta_\nu)}e^{\tau_I^\xi(\theta_I)}\ket{\phi_{ref}^\xi}
\end{eqnarray}
Here $\nu$ represents the combined hole-particle indices of scatterers of embedded subsytem 
$\xi$.
Now only those scatterers are selected and added to the corresponding operator block 
for which the energy difference $\Delta E_{I\nu}^\xi=(E_I^\xi-E_{I\nu}^\xi)$ is greater than 
predefined threshold ($\Delta E_{I\nu}^\xi>\epsilon_2$). It is important to note that the 
ordering of the parent operator blocks remains unchanged in this step, only the dominant 
scatterers are incorporated into their respective blocks. Subsequently, all single excitation 
operators are appended at the final stage. Since the quantum resource requirements associated 
with single excitation operators are relatively minimal, no additional screening criteria are 
imposed to limit their number. The resulting construction yields the final form of the trial 
high-level wavefunction for each embedded subsystem $\xi$.
\begin{eqnarray}\label{frag_f_ansatz}
    \ket{\psi^\xi(\theta)}=\prod_se^{\tau_s^\xi(\theta_s)}...\Bigg[e^{\sigma_\lambda^\xi(\theta_\lambda)}e^{\sigma_\nu^\xi(\theta_\nu)}e^{\tau_K^\xi(\theta_K)}\Bigg]_{\gamma}\\\nonumber ...\Bigg[e^{\tau_J^\xi(\theta_J)}\Bigg]_{\beta}...\Bigg[e^{\sigma_\nu^\xi(\theta_\nu)}e^{\tau_I^\xi(\theta_I)}\Bigg]_{\alpha}...\ket{\phi_{ref}^\xi}.
\end{eqnarray}
Here single excitation operators are represented by $'s'$.
Such a factorized form which is crucial for the 
exactness of the wavefunction is reminiscent of the contracted Schr\"odinger equation 
(CSE)\cite{mazi}. The CSE and its anti-Hermitian variant (ACSE)\cite{mazi2,mazi3} enable direct evaluation of 
energies and two-electron reduced density matrices, offering greater efficiency and 
expressibility.
By optimizing the variational parameters of the constructed trial ansatz, Eq. \ref{frag_f_ansatz}, one obtains the 
correlated high-level wavefunction along with the corresponding minimum energy
for the embedded 
subsystem $\xi$. Repeating this procedure independently for each fragment enables the evaluation of 
the correlated energies and wavefunctions across all subsystems. Here one may note that
the length and structure of the ansatz varies for different embedded subsystems.
Having discussed the building blocks of our theoretical formulation, we present
below a stepwise methodology that leads to the dynamic construction of high-level wavefunction 
of embedded problem and corresponding energy optimization.

\section{Methodology}\label{method}
\begin{enumerate}[a)]
    \item \textbf{Low-Level Description of the Full System:} The mean-field wavefunction of the entire system is obtained using the Restricted Hartree–Fock (RHF) approach (appropriate for closed-shell systems which are considered here). This calculation provides the molecular orbital coefficient matrix along with the total electron count, which serve as the fundamental inputs for subsequent partitioning of the system.
    
    \item \textbf{Fragment Orbital Construction:} The full system is partitioned into multiple user-defined, atom-centered fragments ($A_y$). For each fragment, localized and orthonormalized molecular orbitals are generated using localization techniques
    (here we use the meta-Lowdin localization scheme), forming the corresponding fragment orbital basis..

    \item \textbf{Construction of Entangled Bath Orbitals:} For every fragment, bath orbitals ($B_y$) are identified from the remaining environmental orbitals that exhibit entanglement with the fragment space. This is determined based on eigenvalues lying between 0 and 1 obtained via diagonalizing the environment subblock of one-body reduced density matrix. Orbitals with eigenvalues close to 1 are treated as fully occupied, while those with eigenvalues below $10^{-13}$ are classified as virtual and excluded from the embedding space. The combination of fragment and bath orbitals ($A_y + B_y$) defines the embedded subsystem $\xi$.

    \item \textbf{DMET Macro Cycle:}

    \begin{enumerate}[$\bullet$]
 
    \item \textbf{Embedded Hamiltonian Formation:} For each subsystem $\xi$, an effective Hamiltonian is constructed by projecting the full system Hamiltonian onto the fragment plus bath orbital space. A global chemical potential ($\mu_{gl}$) term is incorporated and adjusted to ensure particle number conservation, thereby enforcing self-consistency across all embedded subsystems. (Eq. \eqref{emb-hamiltonian})

    \item \textbf{Dynamical Design of High-level Wavefunction using Embedded Hamiltonian}
    Starting from the mean-field reference state and the embedded Hamiltonian for each subsystem $\xi$, a correlated high-level wavefunction is constructed by systematically building a compact and dynamic parametrized ansatz through the following procedure:
    \begin{enumerate}
    \renewcommand{\labelenumi}{(\roman{enumi})}
        \item \textbf{One-Parameter Local VQE Micro-cycles:} For each subsystem $\xi$, the dominant two-body excitation operators are identified through single-parameter VQE optimization cycles. Based on their individual energy contributions relative to the mean-field reference, the selected operators are organized into different operator blocks. (Eq. \eqref{one_par})

        \item \textbf{Commutativity-Based Operator Selection for Higher-Order Effects:} For the selected two-body excitation operators within individual operator blocks, suitable scatterers are chosen by analyzing commutativity relations or commonality of the corresponding CSOs. The restriction of CSOs require us to identify the active orbitals
        for each embedding problem.
        This enables the effective incorporation of higher-order (e.g., triple) excitation contributions through nested commutators.

        \item \textbf{Two-Parameter Local VQE Micro-cycles:} The most significant triple excitation effects are further screened via optimizing two-parameter VQE circuits, each composed of one two-body excitation operator and one non-commuting scatterer. 
        This step ensures inclusion of essential correlation effects while maintaining reduced circuit complexity. (Eq. \eqref{two_par})

        \item \textbf{Global VQE Optimization Cycle:} The final ansatz for each subsystem comprising the selected two-body operators, associated scatterers, and all single excitation operators is optimized variationally. The optimization is initialized using parameter values obtained from the preceding local screening cycles to enhance convergence efficiency.
        \begin{eqnarray}
            E^\xi(\theta)=\bra{\psi^\xi(\theta)}\hat{H}^\xi_{emb}(\mu_{gl})\ket{\psi^\xi(\theta)} ; \theta\in \xi
        \end{eqnarray}

    \end{enumerate}
    \end{enumerate}
    \item \textbf{Energy and Electron Number Evaluation for Each Embedded Subsystem:}
    From the correlated high-level wavefunction of each embedded subsystem, the one- and two-particle reduced density matrices (1-RDM and 2-RDM) are evaluated. These quantities are then used to compute the energy of each subsystem. The electron population within a given fragment is obtained as
    $N_e^{A_y}(\mu_{gl})=\sum_{p \in A_y} D_{pp}(\mu_{gl})$,
    where $D_{pp}(\mu_{gl})$ are the diagonal elements of the 1-RDM. The total electron number across all fragments is subsequently determined by summing the individual fragment contributions:
    \begin{eqnarray}
    N_{tot}(\mu_{gl})=\sum_{A_y} N_e^{A_y}(\mu_{gl})
    \end{eqnarray}

    \item \textbf{Self-consistency Check:} The consistency of the electron number is assessed by evaluating the deviation $\Delta N_e = N_{tot}(\mu_{gl}) - N_{act}$, where $N_{act}$ denotes the total number of electrons in the full system. If $\Delta N_e$ falls below a predefined tolerance, the procedure is terminated and the total energy of the system is computed. Otherwise, the global chemical potential $\mu_{\text{gl}}$ is updated, the embedded Hamiltonians are reconstructed accordingly, and the subsequent steps (starting from step d) are repeated until convergence is achieved.
\end{enumerate}    

Below we discuss a few pilot applications to realistic chemical systems that convincingly prove the superiority of dynamic high-level wavefunction construction strategy.

\begin{figure}
    \includegraphics[width=6.0 cm,height=6 cm]{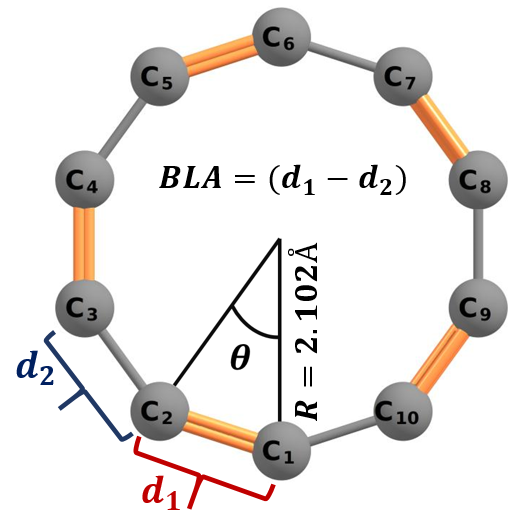}
    \caption{\textbf{A schematic structure of planer $C_{10}$ is presented. The angle between two nearest carbon atoms is defined by $\theta$. The two types of $C-C$ bond lengths
    are presented via $d_1$ and $d_2$, and their difference $(d_1-d_2)$ defines bond length alternation (BLA)}}\label{fig:c10}
\end{figure}

\section{Results and Discussions}\label{results}

\subsection{General Considerations}\label{GC}
In all the simulations, we have used Tangelo\cite{tangelo} which imports one- and 
two-electron integrals from PyScf\cite{pyscf} and created fragment and bath orbitals.
For the calculation of high-level wavefunction, we have integrated Qiskit-nature\cite{qiskit_nature}
in the Tangelo framework. In all the test cases we have used $STO-3G$ basis\cite{sto}
set and employed Jordan-Wigner encoding for fermionic to qubit mapping\cite{JWT}.
In all the test cases, we have employed energy thresholds of $10^{-5}$ and $10^{-7}$
for operator selection in the one-parameter and two-parameter local VQE micro-cycles, 
respectively, for each embedded subsystem. To reduce the number of parameters, we have 
considered two 
specific sets of scattering operator pools.
In the first pool, we include only those two-body scattering operators where the 
transition occurs either from the same spatial orbitals (for $S_h$ type scatterer) or into the same spatial 
orbitals (for $S_p$ type scatterer) of the mean-field reference of the embedded subsystem. This set is referred to as 
the partial-pairing scattering operator pool.
In the second pool, we consider only those scattering operators for which the excitation and 
scattering vertices belong to different spin sectors with respect to the mean-field 
reference of the embedded subsystem. This set is termed the opposite-spin scattering 
operator pool.
For both pools, the corresponding CSOs are restricted to the HOMO and LUMO orbitals of the 
embedded subsystem’s mean-field reference. In the VQE optimization we have used 
SLSQP optimizer (as implemented in SCIPY library)\cite{scipy}.
Furthermore, for the convergence of the DMET 
cycle, we impose a threshold of $\Delta N_e=10^{-5}$ in all test simulations.
In the DMET-cycle we have used Newton-secant root finding method to update the global chemical potential\cite{newton}. 
In the case of ADAPT-VQE\cite{adapt} simulations, we have employed a gradient threshold of 
$10^{-4}$. This implies 
that, during the construction of the high-level wavefunction for each embedded subsystem, 
operators corresponding to the largest gradient are iteratively added to the ansatz until the 
maximum gradient falls below the specified threshold. For these ADAPT-VQE simulations, we have 
used single and double excitation operators pool. 

\begin{figure}
    \includegraphics[width=8.0 cm,height=11.5 cm]{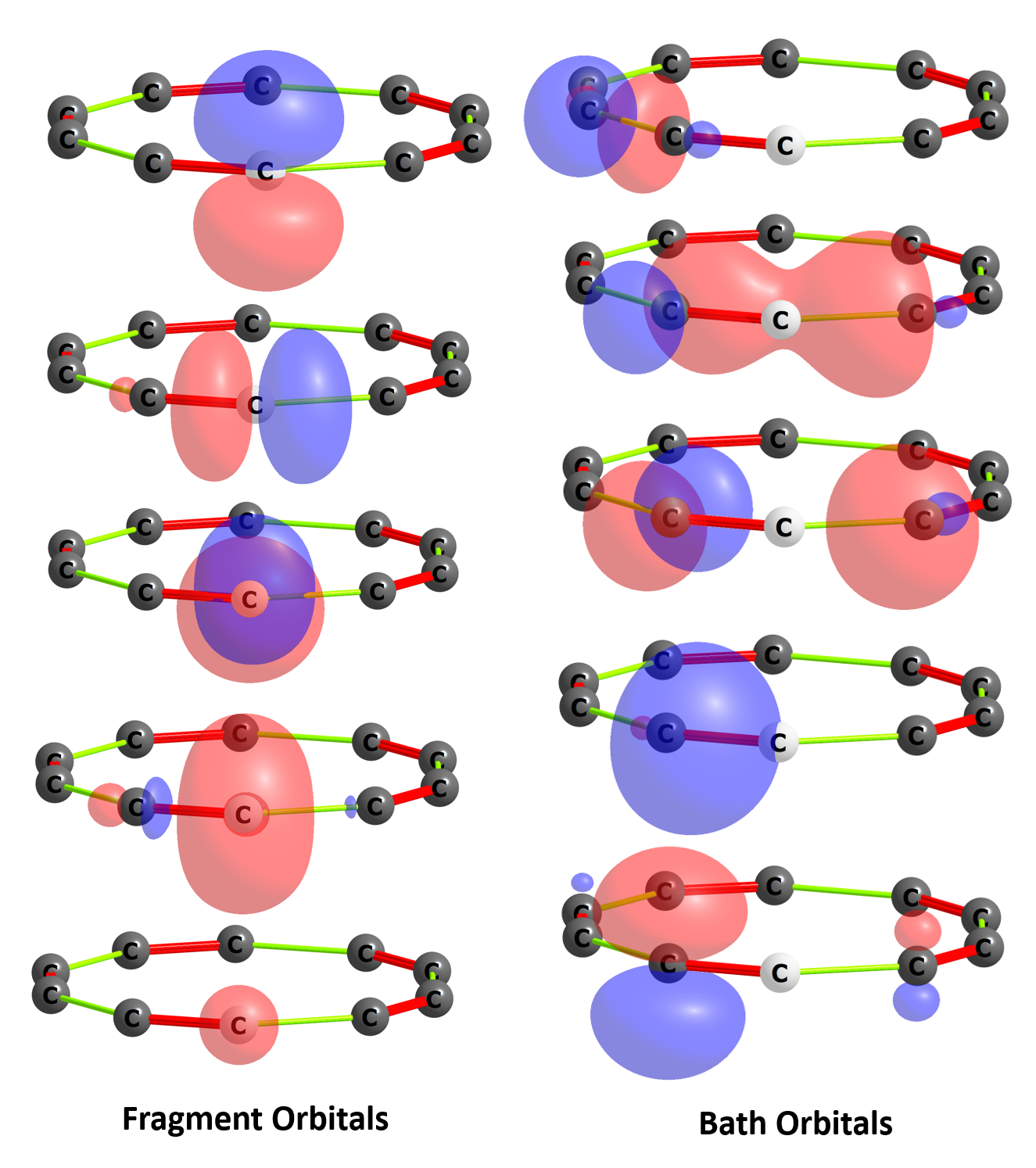}
    \caption{\textbf{The fragment and bath orbitals are presented for $C_{10}$ obtained by partitioning the full system into 10 equivalent C atoms. The orbitals correspond to that carbon atom which is colored white.}}\label{fig:orbitals}
\end{figure}

\begin{figure*}[t]
    \centering
    \includegraphics[width= 17cm, height=15cm]{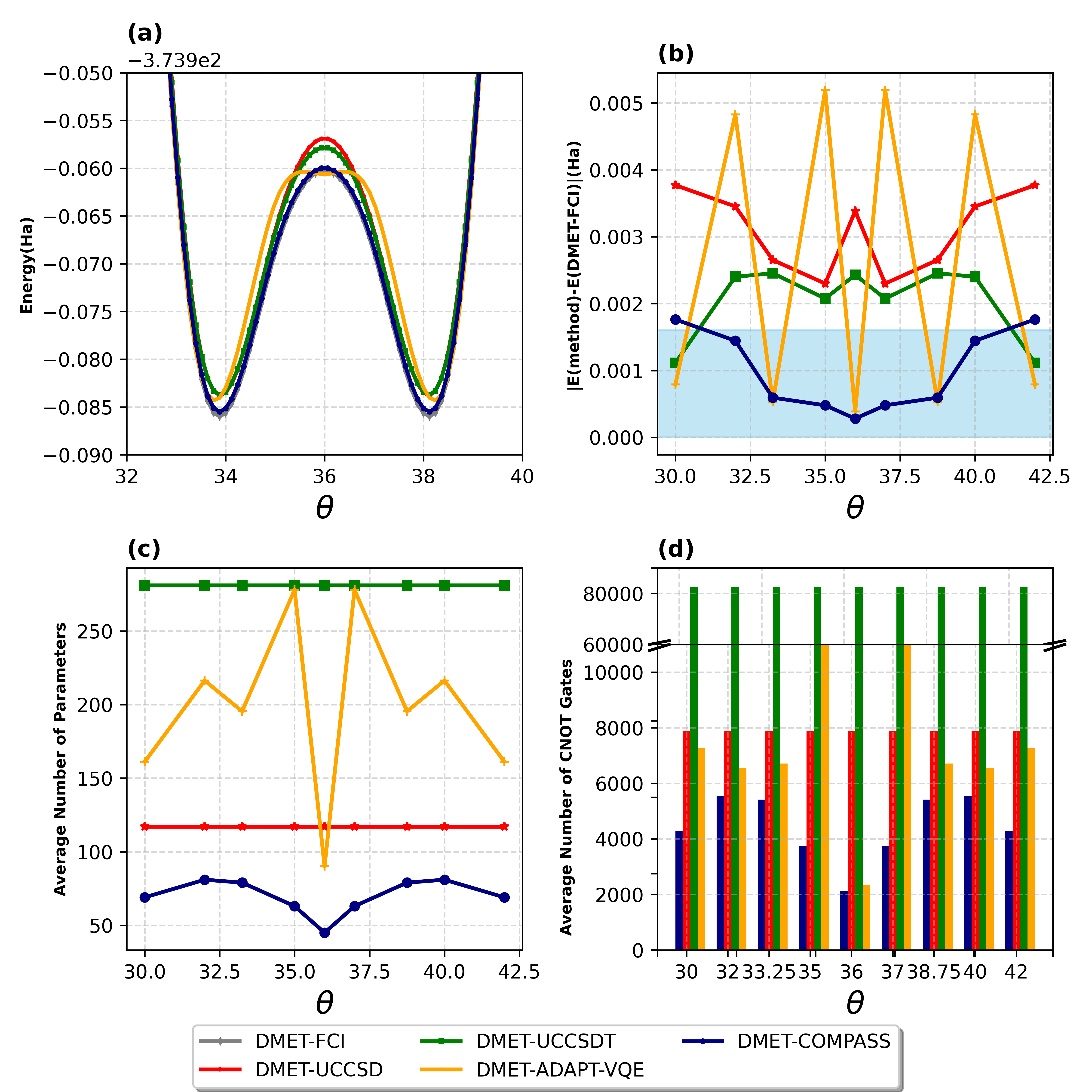}
    \caption{\textbf{The total energy of various methods is presented in (a), while the energy difference from DMET-FCI is presented in (b). The shaded blue region in (b) indicates chemical accuracy with respect to DMET-FCI. In second row, the corresponding average number (across all DMET-cycle) of parameters and CNOT gate counts are shown in (c) and (d), respectively for all methods.}}
    \label{fig:c10_results}
\end{figure*}

\subsection{Potential Energy Profile of Cyclo[10]carbon($C_{10}$):}\

\begin{figure*}[t]
    \centering
    \includegraphics[width= 17cm, height=10cm]{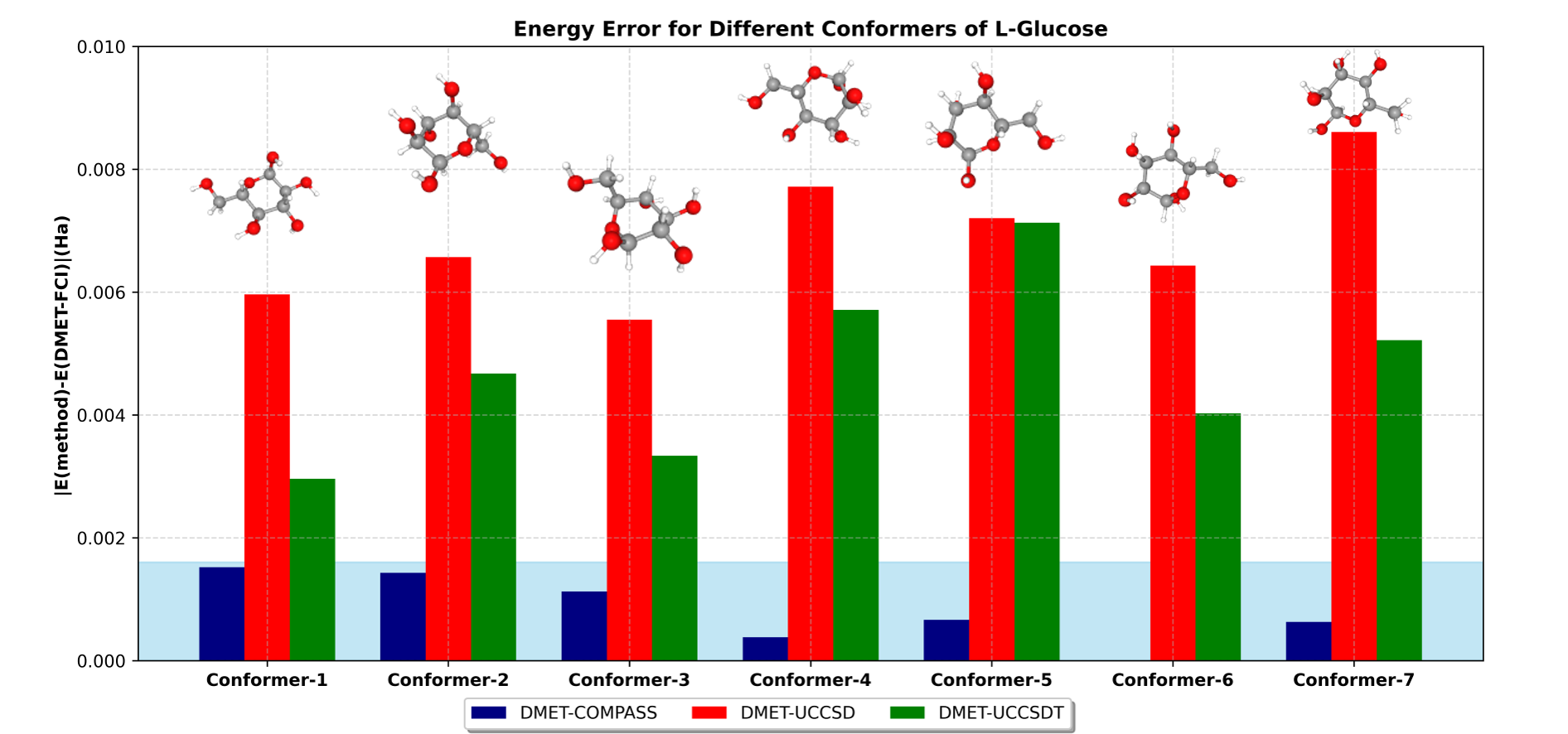}
    \caption{\textbf{Deviation in energy (in $E_h$) from DMET-FCI for different conformers of L-glucose is shown for DMET-UCCSD, DMET-UCCSDT and DMET-COMPASS. The shaded light blue region indicates chemical accuracy with respect to DMET-FCI.}}
    \label{fig:glucose}
\end{figure*}

Following the discovery of carbon allotropes such as fullerenes, carbon nanotubes, and 
graphene, another intriguing class of all-carbon systems cyclo[n]carbons ($C_n$) has 
attracted significant attention due to their unique structural and electronic properties. 
In this work, we 
investigate various geometries of cyclic $C_{10}$ by systematically varying the bond 
angle ($\theta$) between adjacent carbon atoms. All carbon atoms are constrained to lie 
in a single plane. The structure is characterized by two distinct $C-C$ bond 
lengths ($d_1$ and $d_2$), and their difference defines the bond length alternation (BLA: 
$(d_1 - d_2)$). A structural demonstration of the system is presented in Fig. \ref{fig:c10}. 

At $\theta = 36^\circ$, the BLA becomes zero, leading to a cumulenic structure with 
$D_{10h}$ symmetry. For all other values of $\theta$, the system adopts a polyynic 
structure with $D_{5h}$ symmetry and exhibits a non-zero BLA. In our study, the radius 
($R$) of the cyclic $C_{10}$ ring is fixed at 2.102 \AA. The VQE simulation of
full $C_{10}$ system in $STO-3G$ basis requires 100 qubits.

To analyze the system within the embedding framework, we partition the molecule into 10 
fragments using an atom-centered decomposition. The fragment molecular orbitals and 
the corresponding entangled bath orbitals for a carbon fragment of $C_{10}$ are shown in Fig. \ref{fig:orbitals}.
For each fragment, an embedded 
Hamiltonian is constructed, and the corresponding energy is evaluated by preparing 
high-level correlated wavefunctions using the COMPASS-VQE approach, as well as standard 
UCCSD and 
UCCSDT ansatze, within the iterative DMET self-consistency cycle. In this study
the high-level wavefunction for each fragment is calculated in (6,6) active space
keeping the two core occupied and two high line unoccupied orbitals of embedded subsystem 
frozen. Thus for the simulation of high-level wavefunction in quantum computer for each embedded subsystem requires 12 qubits.
The obtained results 
are benchmarked against DMET-FCI calculations within that active space, where the 
high-level wavefunction for each 
embedded subsystem is determined using the exact FCI
method at every step of the global chemical potential optimization. For the DMET-COMPASS
simulation we have used partial-pairing scattering operator pool. (see subsection \ref{GC})

From Fig. \ref{fig:c10_results}, it is evident that the accuracy of DMET-COMPASS is significantly superior to 
that of DMET-UCCSD across the entire range of geometries, and for most configurations it 
even achieves better quantitative agreement than DMET-UCCSDT. Unlike DMET-UCCSD and DMET-
UCCSDT, where the ansatz structure remains fixed, the ansatz in DMET-COMPASS is 
dynamically adapted in each DMET cycle. Therefore, we report the number of parameters and 
CNOT gate counts averaged over all DMET cycles. As shown in Fig. \ref{fig:c10_results}, the average parameter 
count and CNOT gate requirements for DMET-COMPASS are substantially lower than those of 
both DMET-UCCSD and DMET-UCCSDT, highlighting its superior resource efficiency.

We have also evaluated the ground state energies using ADAPT-VQE, a widely used dynamic 
ansatz construction approach, within the DMET framework using the same active space and 
fragmentation scheme as employed in DMET-COMPASS. From Fig.\ref{fig:c10_results} , it is clear that 
while DMET-ADAPT-VQE achieves accuracy comparable to DMET-COMPASS for certain geometries, 
its performance is not consistent across the potential energy profiles. Notably, in all cases, DMET-
ADAPT-VQE requires a significantly larger average number of parameters and CNOT gates 
compared to DMET-COMPASS. This is due to its gradient-based operator 
selection, which often encounters convergence plateaus, necessitating the inclusion of 
additional operators to achieve convergence. 
Furthermore, DMET-ADAPT-VQE typically requires more 
number of DMET cycles to converge the global chemical potential. These observations 
collectively demonstrate the effectiveness and efficiency of the proposed DMET-COMPASS 
approach.

\subsection{Ground State Energy Calculation for Different Conformers of L-Glucose}
Carbohydrate chemistry plays a fundamental role in understanding the structure, 
reactivity, and biological function of sugars, where stereochemistry is a key governing 
factor. Within this framework, L-glucose, the enantiomer of naturally abundant D-glucose, 
provides an ideal model to investigate the effects of stereochemistry on molecular 
properties. In this context, the ground state energy of various conformers of L-glucose may be considered an important factor to determine their relative stability which we decipher using our newly developed algorithm. The molecular geometries for all conformers 
are obtained from PubChem\cite{pubchem_2026_l_glucose}.

Similar to the previous case, the full system is partitioned in an atom-wise manner, which 
significantly reduces the qubit requirement from 144 to at most 20 qubits for the VQE 
simulations. In this setup, each fragment corresponds to an individual H, C, or O atom. For the 
C and O fragments, high-level wavefunctions are computed within a (4,4) active space using 
DMET-COMPASS, DMET-UCCSD, and DMET-UCCSDT approaches. The accuracy of these methods is benchmarked 
against DMET-FCI results evaluated within the same active space. In the DMET-COMPASS simulation
for these test cases we have used the opposite-spin scattering operator pool as previously defined in \ref{GC}.

Although all the DMET based quantum algorithms predict correct relative energy ordering as DMET-FCI
as shown in Fig.\ref{fig:glucose}, DMET-COMPASS exhibits excellent quantitative agreement. In contrast, the deviations observed for DMET-UCCSD and DMET-UCCSDT relative to DMET-
FCI are significantly larger, exceeding the chemical accuracy threshold of 1.6 mHa. A key 
advantage of DMET-COMPASS is its dynamic nature: during each DMET cycle, the ansatz structure 
and length dynamically change in response to variations in the global chemical potential. This is 
in contrast to DMET-UCCSD and DMET-UCCSDT, where the ansatz remains fixed throughout the DMET 
iterations.
For the different L-glucose conformers, the maximum number of parameters (CNOT gates) required 
for the high-level wavefunction calculations is 24 (1016 CNOT gates) in the case of 
DMET-COMPASS. In comparison, DMET-UCCSD and DMET-UCCSDT require 26 (1126 CNOT gates) and 34 (3870 CNOT
gates), respectively. These results clearly demonstrate that DMET-COMPASS is more efficient in 
capturing localized correlation effects within each embedded subsystem while utilizing 
significantly fewer quantum resources compared to DMET-UCCSD and DMET-UCCSDT.

\begin{figure*}[t]
    \centering
    \includegraphics[width= 17cm, height=8cm]{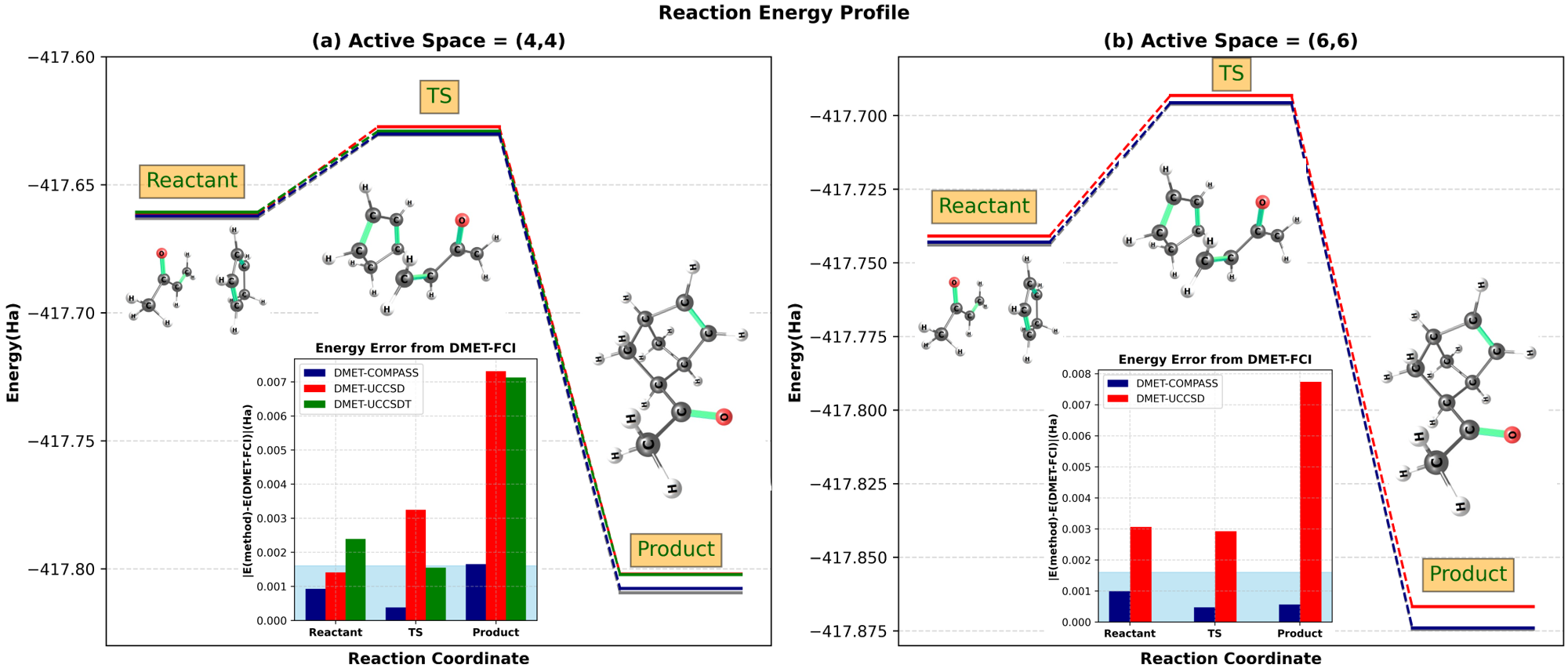}
    \caption{\textbf{Reaction energy profile of (4+2) cycloaddition reaction between cyclopentadiene and  methyl vinyl ketone is presented for DMET-UCCSD, DMET-UCCSDT and DMET-COMPASS.
    The two different plots ((a) and (b)) indicate the different active space dimensions ((4,4) and (6,6) respectively) for high-level wavefunction calculation of the embedded subsystems (for C and O atom fragments). The inset plots
    indicate the corresponding energy deviation from DMET-FCI. The blue shaded region indicates chemical accuracy with respect to DMET-FCI}}
    \label{fig:diels-alder}
\end{figure*}

\subsection{Reaction Energy Profile of Diels–Alder Reaction:}
In this section, we investigate the reaction energy profile of the Diels-Alder reaction, one of 
the most widely utilized and extensively studied transformations in organic chemistry. As a 
representative case, we consider the cycloaddition between methyl vinyl ketone (MVK) and 
cyclopentadiene (CP). The optimized geometries of the reactant, transition state, and product 
employed in this work were obtained from prior DMET-D3 calculations\cite{dmet-d3}. In this benchmarking study
we divided the whole systems (reactant, intermediate and product) into atom-wise
fragment. Each fragment comprises individual atoms (H, C, or O), enabling a significant 
reduction in the effective system size. Through this fragmentation scheme, the original problem 
size of 124 qubits is reduced to embedded subsystems requiring at most 20 qubits.

Within each embedded subsystem (except H fragment) the high-level wavefunction is calculated 
via considering the active space dimension (4,4) and (6,6). In the both situations we have
determined the high-level wavefunction via UCCSD, UCCSDT and COMPASS strategies and compared 
the results with DMET-FCI within the corresponding active space dimensions. In case of DMET-COMPASS we have used partial-pairing scattering operator pool.

From the Fig. \ref{fig:diels-alder}, it is evident that for reactants, intermediates, and products, across both active 
space dimensions, the accuracy of DMET-COMPASS is significantly superior to that of DMET-UCCSD 
and DMET-UCCSDT. This indicates that DMET-COMPASS is more effective in capturing localized 
electron correlation compared to the other two approaches. In particular, in case of the product 
where two new $\sigma$-bonds are formed from the breaking of two $\pi$-bonds during the Diels–
Alder reaction, DMET-COMPASS demonstrates a notably improved description of strong correlation 
effects. Moreover, in all the cases, the DMET-COMPASS energies reach chemical accuracy with 
respect to DMET-FCI calculated in their respective active space dimensions. Importantly, 
for the 
product, which exhibits strong correlation, the high-level wavefunction calculations performed 
in a (6,6) active space across the C and O fragments demonstrate a substantial reduction in 
quantum resources. On average (considering all C and O fragments over all DMET cycles), DMET-
COMPASS requires 4029 CNOT gates. In comparison, the corresponding DMET-UCCSD calculations 
require 7893 CNOT gates, while DMET-UCCSDT demands 84477 CNOT gates. This clearly 
highlights the superior resource efficiency of DMET-COMPASS.

\begin{figure*}[t]
    \includegraphics[width=17.0 cm,height=12 cm]{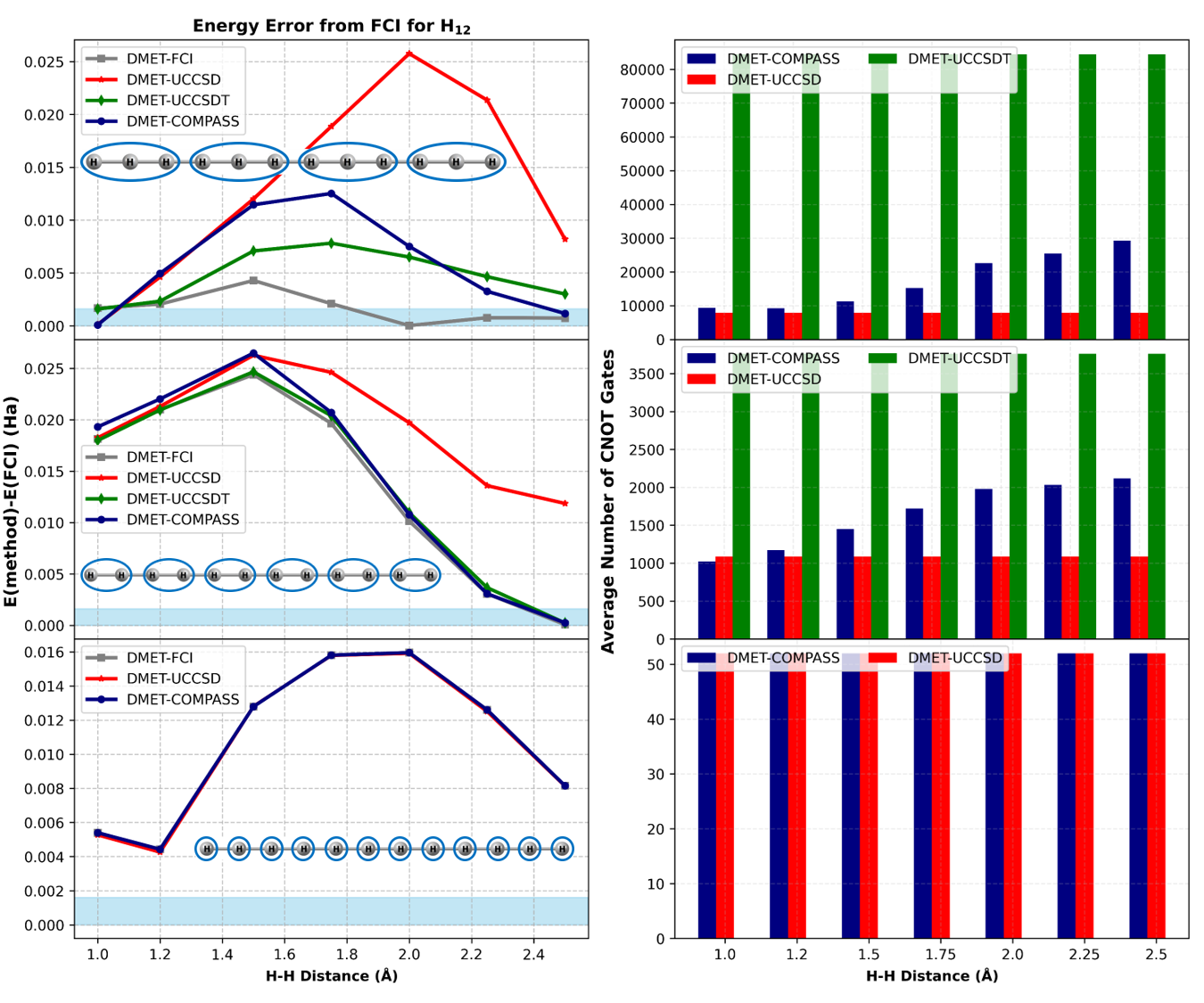}
    \caption{\textbf{Energy errors with respect to canonical FCI are presented for different methods and fragmentation schemes for the linear $H_{12}$ system as a function of the $H$–$H$ bond length. The shaded blue region denotes chemical accuracy relative to FCI. The corresponding average number of CNOT gates (averaged over all fragments and DMET cycles) are shown in the bar plots.}}\label{fig:h12_surface}
\end{figure*}

\subsection{Effects of Different Fragmentation Schemes: Accuracy vs Quantum Efficiency}\label{frag_tech}
In all the previous test cases, we constructed the embedded subsystems by partitioning the full 
system in an atom-wise manner and benchmarked the results of our approach against DMET-FCI. 
However, it is equally important to assess how the various DMET based strategies perform with respect to the exact full
system FCI calculation and how this deviation depends on the chosen fragmentation scheme.
To this end, we consider a moderately sized, strongly correlated system, the linear $H_{12}$ 
chain, for which the exact FCI energy can be computed within the STO-3G basis. In this system, 
all $H$–$H$ bond distances are kept identical, and the potential energy surface is explored by 
varying the bond length from 1.0\AA\ to 2.5\AA.
We adopt three different fragmentation schemes: in the first case, the system is divided into 12 
individual $H$ atoms; in the second case, it is partitioned into 6 fragments, each consisting of 
an $H_2$ unit; and in the third case, the system is divided into 4 fragments, each containing an 
$H_3$ unit. For all fragmentation schemes, we compute the ground state energies using DMET-FCI, 
DMET-COMPASS, and DMET-UCCSD, and compare the results against the canonical exact FCI energy of 
the full system.

It is important to note that in the first fragmentation scheme, where linear $H_{12}$ is 
partitioned into 12 individual H atoms, each embedded subsystem contains 2 electrons in 4 spin 
orbitals. In this case, only one double excitation is possible, with no 
scope for triple or higher-order excitations. Consequently, within the DMET-COMPASS framework, no 
scattering operators are involved. As a result, the performances of DMET-COMPASS and DMET-UCCSD 
are expected to be identical, which is also reflected in Fig. \ref{fig:h12_surface} . However, the results obtained from DMET-based methods throughout
the potential energy surface remain relatively poor for this fragmentation scheme in general.
In the second fragmentation scheme, where each fragment consists of one $H$–$H$ unit, the 
accuracy of DMET-FCI improves significantly, particularly in the strongly correlated regime. 
However, this improvement is not uniform across the entire potential energy surface. In this 
case, both DMET-COMPASS and DMET-UCCSDT exhibit comparable accuracy to the DMET-FCI results 
throughout the surface, whereas DMET-UCCSD shows noticeable deviations, especially at larger bond 
distances as seen from Fig. \ref{fig:h12_surface}. The average number of CNOT gates required for the DMET-COMPASS ansatz is slightly 
higher than that of DMET-UCCSD, yet remains substantially lower than that of DMET-UCCSDT.
Finally, in the third fragmentation scheme, where each fragment comprises an $H$–$H$–$H$ unit, 
the accuracy of DMET-FCI improves considerably across the entire potential energy surface, 
approaching chemical accuracy in most regions. While DMET-UCCSD performs poorly throughout, 
DMET-COMPASS demonstrates strong quantitative agreement with DMET-FCI, particularly in the strongly 
correlated regime. Although minor deviations are observed at intermediate bond distances, its overall 
performance is significantly better than DMET-UCCSD. In terms of quantum resources, the average 
CNOT gate count for DMET-COMPASS is higher than that of DMET-UCCSD but remains substantially 
lower than that required for DMET-UCCSDT. The accuracy and resource efficiency of the 
DMET-COMPASS framework can be further enhanced by appropriately modifying the CSOs involved in the 
scattering operators (currently only the HOMO and LUMO are taken as CSO across all fragmentation
schemes) and by systematically eliminating those scatterers that introduce redundant 
higher-order contributions. Such refinements help to reduce unnecessary circuit complexity 
while retaining the essential correlation effects.
It is particularly important to note that in the final fragmentation scheme, where each embedded 
subsystem requires a 12 qubit VQE simulation within the DMET framework, it is quite possible 
to reach within the chemical accuracy with respect to the canonical FCI energy. In contrast, a 
full simulation of the $H_{12}$ system on a quantum computer would require 24 qubits, which is 
twice the number needed in the DMET-based approach. This clearly demonstrates the superior 
resource efficiency of DMET methodologies for large scale molecular simulations.
In this case, it is observed that increasing the fragment size tends to improve the accuracy of 
DMET-based methods. However, this trend is not universal, and the optimal fragmentation strategy 
is inherently system dependent.

\begin{figure*}[t]
    \includegraphics[width=17.0 cm,height=7 cm]{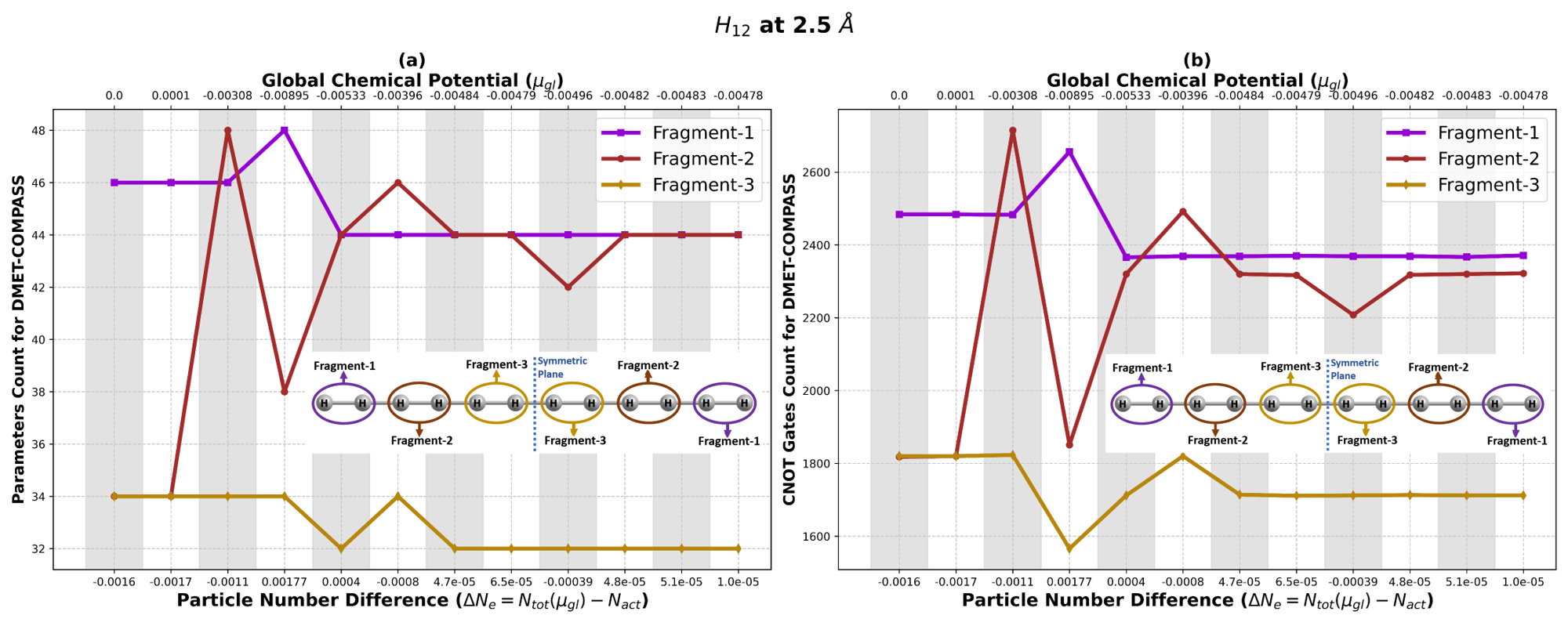}
    \caption{\textbf{Variation in the (a) number of parameters and (b) CNOT gate count for the DMET-COMPASS ansatz as functions of the global chemical potential at each stage of the DMET optimization cycle are presented for different inequivalent fragments of the linear $H_{12}$ system. The $H$–$H$ bond length is fixed at 2.5 \AA. At each cycle, with the variation of global chemical chemical potential, the convergence of the number of particles is also presented.}}\label{fig:parcounts}
\end{figure*}

\subsection{Adaptability of the DMET-COMPASS Ansatz with Respect to Global Chemical Potential}
\label{ansatz_adaptability}
In this subsection, we demonstrate how the DMET-COMPASS ansatz corresponding to each embedded 
subsystem dynamically adapts with the variation of the global chemical potential ($\mu_{gl}$) 
across successive DMET macro cycles. For this purpose, we consider the linear $H_{12}$ system at 
an $H$–$H$ bond length of 2.5 \AA, where the full system is partitioned into six fragments, each 
comprising an $H$–$H$ unit.
As this is a symmetric system with respect to the plane consisting the point of inversion, 
three different types of fragment units are present here. 

For each of these fragment types, we analyze how the DMET-COMPASS ansatz length changes over 
different DMET macro-cycles. In Fig. \ref{fig:parcounts}, we present the variation of the number of variational 
parameters along with the corresponding CNOT gate counts as functions of the global chemical 
potential and the particle number difference at each macro cycle. It is evident from Fig. \ref{fig:parcounts}, that 
different fragments require different ansatz sizes within the DMET-COMPASS framework. More 
importantly, even for a given fragment, the ansatz length varies from one DMET macro cycle to 
another. This behavior highlights that, as the global chemical potential is updated at each macro-cycle, the corresponding embedded Hamiltonians are 
modified. As a result, the DMET-COMPASS ansatz
auto-adjusts to accurately capture the correlation effects. Therefore, DMET-COMPASS inherently 
involves a coupled optimization of the global chemical potential and the ansatz structure. This 
coupling provides enhanced flexibility, enabling the ansatz to adapt consistently with the changes 
in the global chemical potential. Furthermore, Fig. \ref{fig:parcounts} shows that the variation in ansatz length is 
more pronounced during the initial DMET cycles and gradually stabilizes as convergence is 
approached. It is also important to note that, although two different fragments may require 
the same number of parameters (as seen in Fig. \ref{fig:parcounts}(a)), their corresponding CNOT gate counts (Fig. \ref{fig:parcounts} 
(b)) can differ. This indicates that distinct operator sets are dynamically picked up to form 
the ansatz in different fragments.
While in most cases DMET-COMPASS and DMET-UCCSD take similar number of DMET cycles 
to converge the global chemical potential, there may exist some cases (like the one presented
in Fig. \ref{fig:parcounts}), where DMET-COMPASS takes more cycles. However, in such cases
DMET-COMPASS keeps adjusting the ansatz length and ordering for each embedded subsystem over more DMET cycles to arrive at a 
significantly better energy values than DMET-UCCSD.

\section{Conclusions}
Due to the limitation of fault-tolerant qubits and the limited fidelity of quantum gates, 
performing accurate realistic chemical simulations remains beyond the reach of present day 
quantum hardware. To address this challenge, partition-based embedding approaches 
provide an effective 
route by reducing the size of the full system while preserving accuracy through the evaluation 
of localized correlation energies within individual fragments. The overall performance of such 
embedding frameworks can be enhanced in two primary ways: first, by refining the embedding 
scheme to achieve a more accurate representation of the embedded subsystems, and second, by 
improving the quality of the correlated high-level wavefunction so that it can more effectively 
capture the localized correlation effects and use the high-level wavefunction to self-
consistently update the embedding Hamiltonian.

In this work, we employ the latter strategy within the DMET framework to enhance the 
description of ground state of chemically relevant systems and processes. 
The procedure relies on the capturing the localized correlation energy via the
construction of dynamic high-level wavefunction of multiple embedding problems through a 
systematic partitioning of the full system. For each embedded Hamiltonian, a dynamic ansatz is 
generated via commutativity based local operator selection and energy driven operator screening 
procedure tailored to the corresponding subsystem. During every DMET iteration, these subsystem
specific high-level wavefunctions are used to update the global chemical potential, which in 
turn refines the embedding Hamiltonians.
Unlike any static structured ansatz for the construction of the high-level wavefunction,
working with a dynamic ansatz in the DMET framework warrants the coupled optimization
of the global chemical potential (thereby the embedding Hamiltonian) and the 
ansatz structure for each of the embedding problems. 
This introduces additional flexibility in the ansatz design and operator ordering. 
Our numerical applications involving strong correlation  
suggest that such a strategy 
of dynamic ansatz construction within an embedding problem attains very high accuracy
for important chemical problems over
more conventional ansatze with significantly lower quantum gate utilization.

Furthermore, we investigate the improvement in total system energy within the DMET framework as 
a function of fragment size for a strongly correlated linear $H_{12}$ chain. The results 
highlight the critical role of fragmentation scheme in capturing correlation effects, showing that 
increasing fragment size systematically enhances the recovered correlation energy and 
approaches chemical accuracy relative to full system. Moreover, it is demonstrated how
the ansatz adjusts itself for each embedding problem at each step of the DMET cycle as
the global chemical potential (or equivalently embedding Hamiltonian) is updated, 
enabling an accurate representation of correlation effects while maintaining low quantum resource 
requirements.
However, the effectiveness 
of a given partitioning scheme remains system dependent, emphasizing the importance of 
designing optimal fragment-bath subsystem which is an open direction in embedding 
methodologies. 
Additional improvements in both accuracy and resource efficiency of the high-level solver in 
DMET-COMPASS can be achieved through careful tuning of screening thresholds, elimination of 
redundant higher-order excitations, and reduction of measurement overhead via pauli term 
grouping in embedded Hamiltonians\cite{scale,scale2,scale3}. For practical implementations on quantum hardware, 
integration of the high-level solver with proper quantum error mitigation techniques 
is necessary to 
further improve the accuracy\cite{rem,neural,gnm}. Overall, the combination of dynamic, resource-efficient ansatz 
construction within an advanced embedding framework presents a promising route toward accurate and scalable simulations of strongly correlated, classically intractable chemical systems.



\section*{Conflict of Interest:}
The authors have no conflict of interests to disclose.

\section*{DATA AVAILABILITY}
The numerical data that support the findings of this study are
available from the corresponding author upon 
reasonable request.

\section*{Acknowledgment}
DM and RM acknowledge the fruitful discussion with the team of Qclairvoyance Quantum Labs Pvt. 
Ltd. DM thanks Prime Minister's Research Fellowship (PMRF), Government of India for his 
research fellowship. R.M. acknowledges the financial support from the Anusandhan National 
Research Foundation (ANRF) (erstwhile SERB), Government of India (Grant Number: 
MTR/2023/001306).


\bibliography{literature}

\end{document}